\documentclass[preprint,12pt]{aastex}

\newcommand{\eqref}[1]{(\ref{#1})}

\shorttitle{Radiation-Hydrodynamics of Atmospheres}
\shortauthors{Menou and Rauscher}

\begin{document}

\newcommand\bb[1] {   \mbox{\boldmath{$#1$}}  }

\newcommand\del{\bb{\nabla}}
\newcommand\bcdot{\bb{\cdot}}
\newcommand\btimes{\bb{\times}}
\newcommand\vv{\bb{v}}
\newcommand\B{\bb{B}}
\newcommand\BV{Brunt-V\"ais\"al\"a\ }
\newcommand\iw{ i \omega }
\newcommand\kva{ \bb{k\cdot v_A}  }
\newcommand\kb{ \bb{k\cdot b}  }
\newcommand\kkz { \left( \frac{k}{k_Z}\right)^2\>}

    \def\dd{\partial}
    \def\tilde{\widetilde}
    \def\etal{et al.}
    \def\eg{e.g. }
    \def\etc{{\it etc.}}
    \def\ie{i.e.}
    \def\beq{ \begin{equation} }
    \def\eeq{ \end{equation} }
    \def\spose#1{\hbox to 0pt{#1\hss}} 
    \def\ltsim{\mathrel{\spose{\lower.5ex\hbox{$\mathchar"218$}}
         \raise.4ex\hbox{$\mathchar"13C$}}}

\def\tilde{\widetilde}

\newcommand{\schwz}{ {\dd  \ln P\rho ^{-\gamma} \over \dd Z}}
\newcommand{\schwR} { {\dd  \ln P\rho ^{-\gamma} \over \dd R} }

\newcommand{\balbz}{ {\dd  \ln T \over \dd Z}}
\newcommand{\balbR} { {\dd  \ln T \over \dd R} }

\title{Radiation-Hydrodynamics of Hot Jupiter Atmospheres}

\author{Kristen Menou \& Emily Rauscher
  \\ \textit{Department of Astronomy, Columbia University,
  \\ 550 West 120th Street, New York, NY 10027, USA}}

\begin{abstract}
Radiative transfer in planetary atmospheres is usually treated in the
static limit, i.e., neglecting atmospheric motions. We argue that hot
Jupiter atmospheres, with possibly fast (sonic) wind speeds, may
require a more strongly coupled treatment, formally in the regime of
radiation-hydrodynamics. To lowest order in $v/c$, relativistic
Doppler shifts distort line profiles along optical paths with finite
wind velocity gradients. This leads to flow-dependent deviations in
the effective emission and absorption properties of the atmospheric
medium. Evaluating the overall impact of these distortions on the
radiative structure of a dynamic atmosphere is non-trivial. We present
transmissivity and systematic equivalent width excess calculations
which suggest possibly important consequences for radiation transport
in hot Jupiter atmospheres. If winds are fast and bulk Doppler shifts
are indeed important for the global radiative balance, accurate
modeling and reliable data interpretation for hot Jupiter atmospheres
may prove challenging: it would involve anisotropic and dynamic
radiative transfer in a coupled radiation-hydrodynamical flow. On the
bright side, it would also imply that the emergent properties of hot
Jupiter atmospheres are more direct tracers of their atmospheric flows
than is the case for Solar System planets. Radiation-hydrodynamics may
also influence radiative transfer in other classes of hot exoplanetary
atmospheres with fast winds.
\end{abstract}


\section{Introduction} \label{S:intro}

In recent years, it has become possible to remotely observe the
atmospheres of some exoplanets found in close-in orbits around nearby
stars. Many of these observations, which include eclipse, phase curve,
and transit photometric or spectroscopic measurements, have focused on
the specific class of exoplanets known as hot Jupiters
\citep{Charbonneau2002,Charbonneau2005,Deming2005,Charbonneau2008,Knutson2008,Knutson2009b,Harrington2006,Harrington2007,Agol2009,Cowan2007,Knutson2007,Knutson2009a,Tinetti2007,Pont2008,Sing2008,Swain2008a,Swain2008b,Grillmair2007,Richardson2007}. These
gaseous giant planets have high atmospheric temperatures, from the
close proximity to their parent stars, slow spin rotation rates, as
inferred from their expected state of tidal synchronization, and they
are thus subject to an unusual steady pattern of hemispheric
insolation, with permanent day and night sides
\citep[e.g.,][]{Seager2005,SMC08,SCM10}.

A variety of diagnostics about physical conditions in the atmospheres
of a few specific hot Jupiters have been presented in the literature,
from the presence of high altitude haze \citep[e.g.,][]{Pont2008} to
the existence of vertical temperature inversions
\citep["stratospheres", e.g.,][]{Burrows2008,Fortney2008}, as well as
constraints on the chemical abundances of radiatively active species
\citep[e.g.,][]{Charbonneau2002,Tinetti2007,Sing2008,Swain2008b}. In
most cases, these interpretations rely on one-dimensional,
steady-state radiative transfer models describing in great detail the
atmospheric structure that is expected at radiative and chemical
equilibrium. On the other hand, these models usually consider only
globally averaged atmospheric properties and they often ignore the
role of horizontal heat transport by advection on the energetic
balance of the modeled atmospheres. Such shortcomings of radiative
transfer models could eventually limit our ability to reliably infer
the physical conditions present in remotely observed exoplanetary
atmospheres.

Planetary atmospheres which satisfy global radiative equilibrium are
generally not in local radiative equilibrium, that is along a specific
vertical column, because of finite contributions from horizontal heat
fluxes. The role of atmospheric circulation in shaping the structure
and properties of hot Jupiter atmospheres has been recognized as an
increasingly important ingredient over the last few years
\citep[e.g.,][]{Fortney2006} and several groups have been developing
multi-dimensional atmospheric models to address this issue more
explicitly
\citep{SG02,Cho2003,Cho2008,Burkert2005,CS05,CS06,Langton2007,DobbsDixon2008,S08,S09,MR08,RauMen09}.

Even in a three-dimensional, time-dependent atmospheric model of the
type known as "General Circulation Model" (GCM), radiative transfer is
traditionally treated in the static approximation. In this limit, the
coupling between the atmospheric flow and the radiative heat transport
enters only via the energy equation for the moving atmospheric fluid,
so that this coupling is purely thermodynamic in nature. The radiative
transfer component of the problem, which is needed to continuously
evaluate the net diabatic heating or cooling rate of the gas in
motion, is solved independently of the atmospheric motions themselves,
as if the atmosphere were actually static.

Here, we argue that this level of thermodynamic coupling, which is a
standard approximation for Solar System planetary atmospheres, may be
insufficient to describe accurately the nature of energy transport in
hot Jupiter atmospheres. A more strongly coupled treatment known as
radiation-hydrodynamics, in which flow velocities enter the radiation
transport problem explicitly, may be required to describe radiative
transfer in hot Jupiter atmospheres with fast, sonic or transonic wind
speeds.

The remainder of this paper is organized as follows. In
\S\ref{S:dynatm}, we outline the general formalism for radiation
transport in a dynamic atmosphere. We motivate the relevance of this
radiation-hydrodynamics regime for hot Jupiter atmospheres with fast
winds in \S\ref{S:HotJups}. We discuss the potential magnitude of
deviations from the static limit for the global radiative balance of
hot Jupiter atmospheres in \S\ref{S:EW} and we conclude in
\S\ref{S:impl}.

\section{Radiative Transfer in Dynamic Atmospheres} \label{S:dynatm}

Radiation-hydrodynamics requires a fully relativistic treatment to
properly account for Doppler shifts and aberration effects
\citep{Mihalas,Castor04}. The time-dependent radiation transport
equation in the fluid comoving frame, to leading order in the ratio of
the fluid velocity to the speed of light, $v/c$, was derived by
\citet[][see also \citealt{HSpiegel76}]{Buchler83}. The breakdown of
terms in this rather general equation, which may or may not be
discarded depending on the nature of the physical problem at hand, is
a delicate matter \citep{Castor04,Mihalas}. In the limit of slow fluid
motions ($v/c \ll 1$) and thus fast light-transit times, however,
there is general agreement that the leading order effect of fluid
motions is simply to induce Doppler shifts in the instantaneous
radiation field seen by the fluid
\citep[e.g.,][]{Rybicki70,RybHum78}.\footnote{Effects due to
relativistic aberration are thought to be significant only at higher
order in $v/c$ \citep{Castor04}.} This results in modified absorption
and emission coefficients for radiation-matter interaction,
particularly when these interactions are dominated by line
opacities. In the remainder of this work, we focus our discussion on
line opacities, which are expected to dominate in hot Jupiter
atmospheres, and neglect continuum sources of opacities.

Much of the astrophysical literature on radiation-hydrodynamics is
focused on the radiative transfer of radially expanding stellar
winds. We start our discussion closely following the formalism of
\citet{RybHum78}. To lowest order in $v/c$, the radiative transfer
equation in a moving medium can be written

\begin{equation} \label{eq:one}
\bb{n} \bcdot \bb{\nabla} I(\bb{r},\bb{n},\nu)= -k_{\rm tot}(\bb{r}) \Phi\left[ (\nu - \nu_0) + \frac{\nu_0}{c}\bb{n} \bcdot \bb{v(r)} \right] [I(\bb{r},\bb{n},\nu)-S(\bb{r},\bb{n},\nu)],
\end{equation} 
where $I(\bb{r},\bb{n},\nu)$ is the monochromatic intensity (per unit
solid angle) at frequency $\nu$ and spatial location $\bb{r}$ in the
direction defined by the unit vector $\bb{n}$, $S$ is the source
function, $\bb{v(r)}$ is the material velocity field, $k_{\rm tot}$ is
the line integrated opacity, $\nu_0$ is the line central frequency and
$\Phi(\nu)$ is the line profile function, normalized to unity. The
standard radiative transfer equation in the static limit is recovered
by forcing $\bb{v(r)}$ to zero.

Equation~(\ref{eq:one}) expresses the fact that any projected velocity
gradient along an optical path in the direction $\bb{n}$ leads to a
shift in the line profile, and thus a change in the effective
absorption coefficient of the medium, in proportion to the Doppler
term $\bb{n} \bcdot \bb{v(r)} \nu_0 / c$ entering the line function
$\Phi$. By contrast, a uniform velocity field leads to a more trivial,
uniform frequency shift. In the context of stellar wind theory,
equation~(\ref{eq:one}) is typically solved using the approximate
``Sobolev method,'' in the limit of strongly supersonic gradient flow
motions, for radially expanding shells of material
\citep[e.g.,][]{RybHum78,Castor04}. This is not the limit of interest
for hot Jupiter atmospheres.

The literature on radiative transfer in static atmospheres is vast and
the reader is directed to the monographs by \citet{GoodyYung89},
\citet{ThomasStamnes02} and \citet{Liou02} for detailed accounts on
this subject. To make the connection with that work, we now introduce
the plane-parallel atmosphere approximation, using $z$ as the vertical
coordinate in the atmosphere. In addition, we introduce the zenith
angle $\theta$ (away from the vertical axis, pointing upward) and the
azimuthal angle $\phi$ (in the plane normal to the vertical). It is
convenient to introduce the quantity $ \mu = \cos \theta$, which
ranges from $+1$ going up to $-1$ going down in the plane-parallel
atmosphere. We further assume that local thermodynamic equilibrium is
satisfied and we neglect scattering, as is generally justified for the
thermal portion of the atmospheric radiation spectrum.\footnote{It is
customary in the treatment of atmospheric radiation to separate the
short-wavelength component of the spectrum, which is related to
beam-like insolation in the presence of scattering (typically in the
optical), from the long--wavelength component, which is related to
diffuse thermal emission from the various atmospheric layers
(typically in the infrared). We follow this convention here, even
though the two spectral components are not as clearly separated for
hot Jupiters as they are for ``cool'' Solar System planets. For
simplicity, we focus our entire discussion on the long--wavelength
thermal component and postpone a discussion of possible complications
with the short-wavelength portion of the radiation spectrum until
\S\ref{S:impl}.}  The source function $S$ then reduces to the Planck
function, $B_\nu (T(z))=B_\nu (z)$. In that case, the formal solution
to the radiative transfer equation for the monochromatic intensity at
level $z_0$ in the direction $(\mu, \phi)$ can be written
\citep{GoodyYung89,ThomasStamnes02,Liou02}:

\begin{equation} \label{eq:two}
I_\nu(z_0; \mu, \phi)= I_\nu(0; \mu, \phi) {\cal T}_\nu(0,z_0; \mu, \phi) + \int_{0}^{z_0} B_\nu(z; \mu, \phi) \frac{d {\cal T}_\nu(z,z_0; \mu, \phi)}{dz}dz,
\end{equation}
where the monochromatic notation has been changed to $I_\nu$ and $z=0$
corresponds to a bounding region for the atmosphere (e.g., the bulk
interior), with a one-sided radiative boundary condition specified as
$I_\nu(0; \mu, \phi)$. The first term on the RHS of eq.~(\ref{eq:two})
describes the cumulative absorption of $I_\nu(0; \mu, \phi)$ from
$z=0$ to $z_0$ along the optical path, while the second term describes
the cumulative Planck emission, and its absorption, integrated over
all layers along the optical path, from $z=0$ to $z_0$. The
monochromatic transmissivity

\begin{equation} \label{eq:three}
{\cal T}_\nu(z,z_0; \mu, \phi) \equiv \exp \left[ - \int_z^{z_0} k_\nu(z'; \mu, \phi) \rho(z'; \mu, \phi) dz'\right]
\end{equation}
encapsulates the absorption (and emission) properties of the
medium. It is the integral of the mass absorption coefficient $k_\nu$
weighted by the amount of absorber (with mass density $\rho$), between
vertical levels $z$ and $z_0$, along the optical path defined by the
direction $(\mu, \phi$).

According to equation~(\ref{eq:one}), bulk Doppler shifts in a dynamic
atmosphere modify the radiative transfer by changing the monochromatic
absorption coefficient (unit of cross section per unit mass),

\begin{equation} \label{eq:four}
k_\nu(z'; \mu, \phi)= k_{\rm tot}(\bb{r}) \Phi\left[ (\nu - \nu_0) +
\frac{\nu_0}{c}\bb{n} \bcdot \bb{v(r)} \right],
\end{equation} 
or equivalently by changing the monochromatic transmissivities,
${\cal T}_\nu(z,z_0; \mu, \phi)$, in equation~(\ref{eq:two}). If bulk
Doppler shifts are important and the material velocity field is
anisotropic, so is the radiative transfer.

By property of isotropy of the Planck source function, a large
fraction of the thermal atmospheric radiation field is carried along
rays which are significantly slanted relative to the vertical, even
though net flux exchange between atmospheric layers occurs
vertically. The next step in deriving vertical flux equations for the
atmospheric radiation problem typically involves separate angular
integrations for the ascending ($\mu >0$) and descending ($\mu <0$)
fluxes \citep{GoodyYung89,ThomasStamnes02,Liou02}, of the type

\begin{equation}
F_\nu^\pm = \int_0^{2 \pi} \int_0^{\pm 1} I_\nu (\mu, \phi) \mu d\mu d\phi.
\end{equation}

To account for the phase space available at slanted angles, it is
customary in ``two-stream'' formulations to replace all angular
integrals over transmissivities by a single average transmissivity
value for a characteristic zenith angle, $\mu_0= \cos \theta_0$:

\begin{eqnarray}
\int_0^{2 \pi} \int_0^{\pm 1} \exp \left[ - \int_z^{z_0}   k_\nu(z';
  \mu, \phi) \rho(z'; \mu, \phi) dz'\right] \mu d\mu d\phi \nonumber \\  \simeq \pi \exp \left[ - \frac{1}{\mu_0}
  \int_z^{z_0} k_\nu(z'; \mu_0) \rho(z'; \mu_0) dz'\right] \equiv \bar{{\cal T}}_\nu(z,z_0),
\end{eqnarray}
where $\bar{{\cal T}}_\nu$ is the hemispherically-averaged ``diffusive
transmissivity.'' This approximation has been shown to yield errors
$\la 1.5 \%$ in typical applications for static atmospheres, when a
diffusivity factor $1/ \mu_0 = 1.66$ is adopted
\citep{RodgWal66,GoodyYung89,ThomasStamnes02,Liou02}, which
corresponds to a zenith angle $\theta_0 = 53$~deg. This relatively
large value of the effective zenith angle illustrates well the large
phase space available for thermal radiation at slanted angles. It
indicates that the projected wind velocity along a typical optical
path for thermal atmospheric radiation can be a significant fraction
of the full horizontal wind speed,\footnote{Large scale motions are
  predominantly horizontal in an atmosphere.} even though net
radiative exchanges between the atmospheric layers occur in the
vertical.

\section{Hot Jupiter Atmospheres} \label{S:HotJups}

To help us focus our discussion further, we now turn to issues
specific to hot Jupiter atmospheres. 

\subsection{Projected Wind Velocities}

Over the last few years, various circulation models have indicated
that wind speeds could reach sonic or even supersonic values in the
upper atmospheres of hot Jupiters
\citep{CS05,CS06,DobbsDixon2008,S08,S09,RauMen09}. Here, for
concreteness, we use the specific hot Jupiter model described in
\citet{RauMen09} to evaluate the magnitude of projected velocity
gradients along representative optical paths for thermal radiation in
a dynamic hot Jupiter atmosphere.

A specific optical path is defined by a starting location at the
bottom of the model atmosphere and by a unit vector $\bb{n}$ which
defines the path orientation. We calculate the projected wind velocity
along this path throughout the entire model atmosphere as

\begin{equation}
V_{\rm proj}= \bb{n \bcdot V_{\rm h}} = \bb{n_{\rm h} \bcdot  V_{\rm h}} \sin \theta_0 , 
\end{equation}
where $\bb{V_{\rm h}}$ is the wind horizontal velocity vector
according to the circulation model, $\bb{n_{\rm h}}$ is the horizontal
unit vector projected on the sphere (i.e., defining the north/south
and east/west directions) and the representative zenith angle
$\theta_0 = 53$~deg is uniformly adopted in all our calculations.  A
detailed calculation would require us to take into account the
three-dimensional geometry of the problem, with varying pressure
levels in each of the model vertical columns crossed by the slanted
optical path. Rather than performing delicate three-dimensional
interpolations between various model columns, we use the profile of
wind velocities in a single vertical column. That is, we use values of
$\bb{V_{\rm h}}$ as if the optical path were exactly vertical, even
though the calculation assumes a zenith angle $\theta_0 = 53$~deg and
various azimuthal orientations for the projection.  While this
approximation clearly emphasizes vertical velocity gradients over
horizontal ones, it still captures representative changes in
$\bb{V_{\rm h}}$, both in magnitude and direction, along the selected
optical path (with a specific azimuthal orientation). It should thus
be sufficient to evaluate the typical magnitude of projected velocity
gradients along representative optical paths in the model atmosphere.

\begin{figure}[ht]
\vspace{-0.5cm}
\includegraphics[angle=90,scale=0.32]{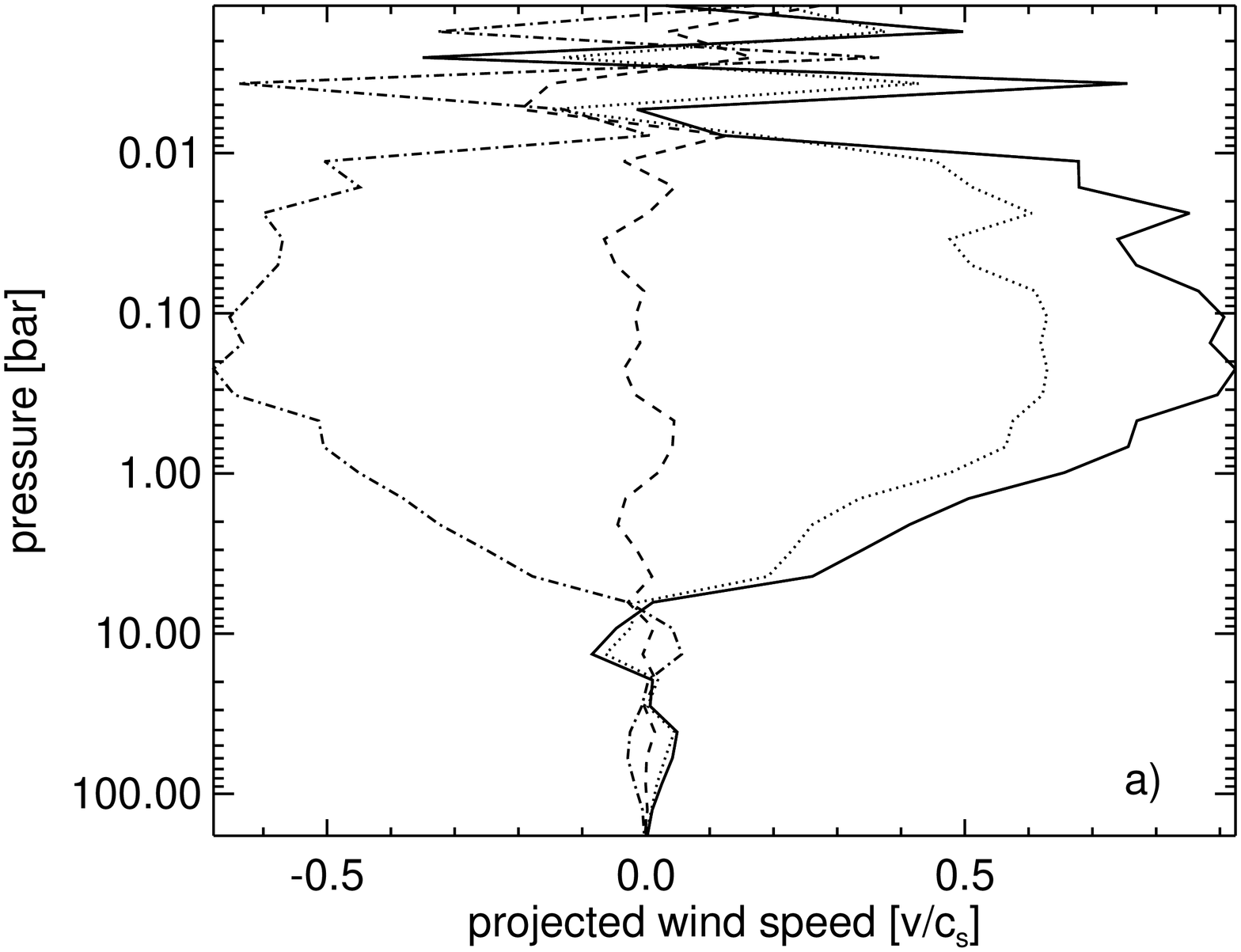}
\includegraphics[angle=90,scale=0.32]{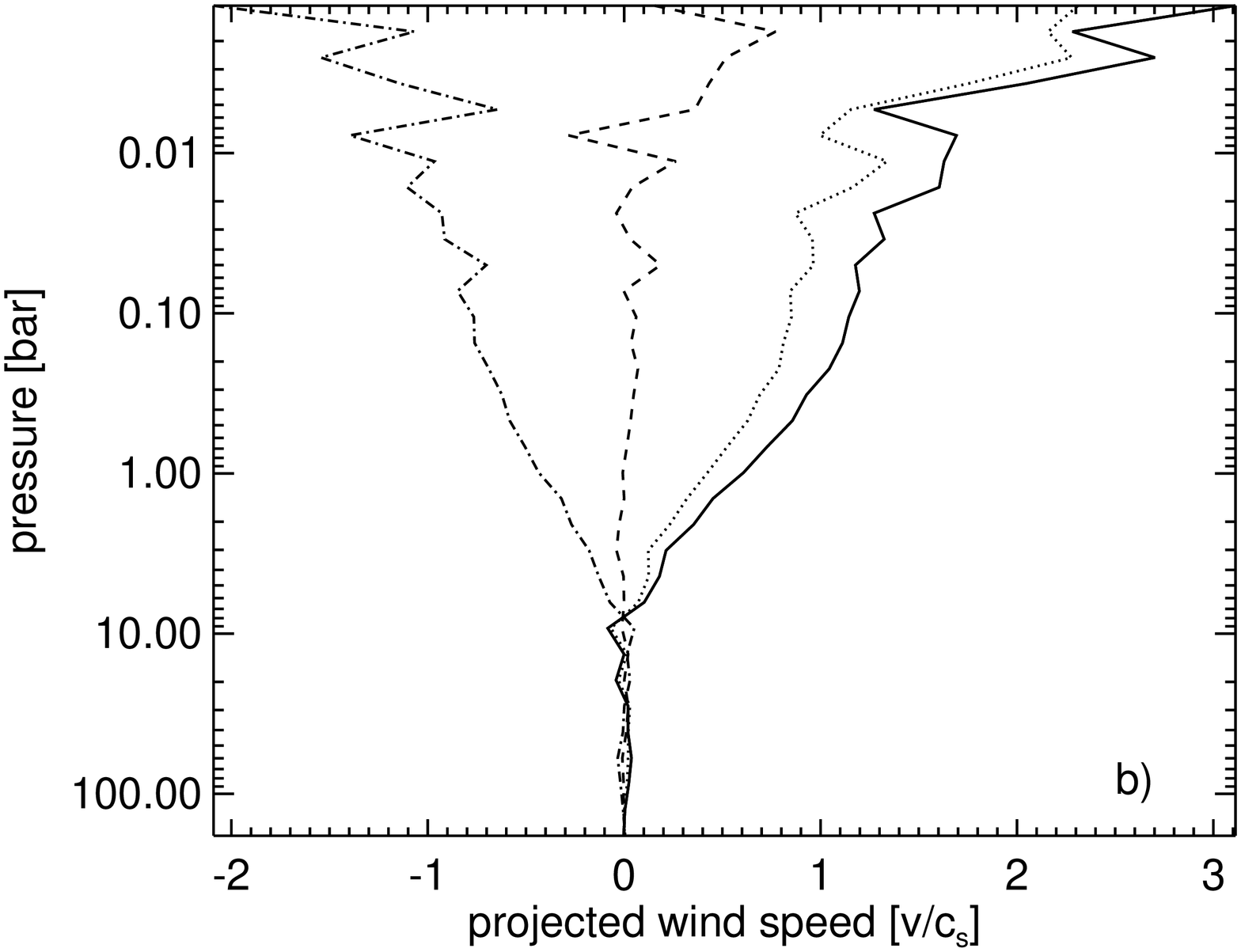} 
\includegraphics[angle=90,scale=0.32]{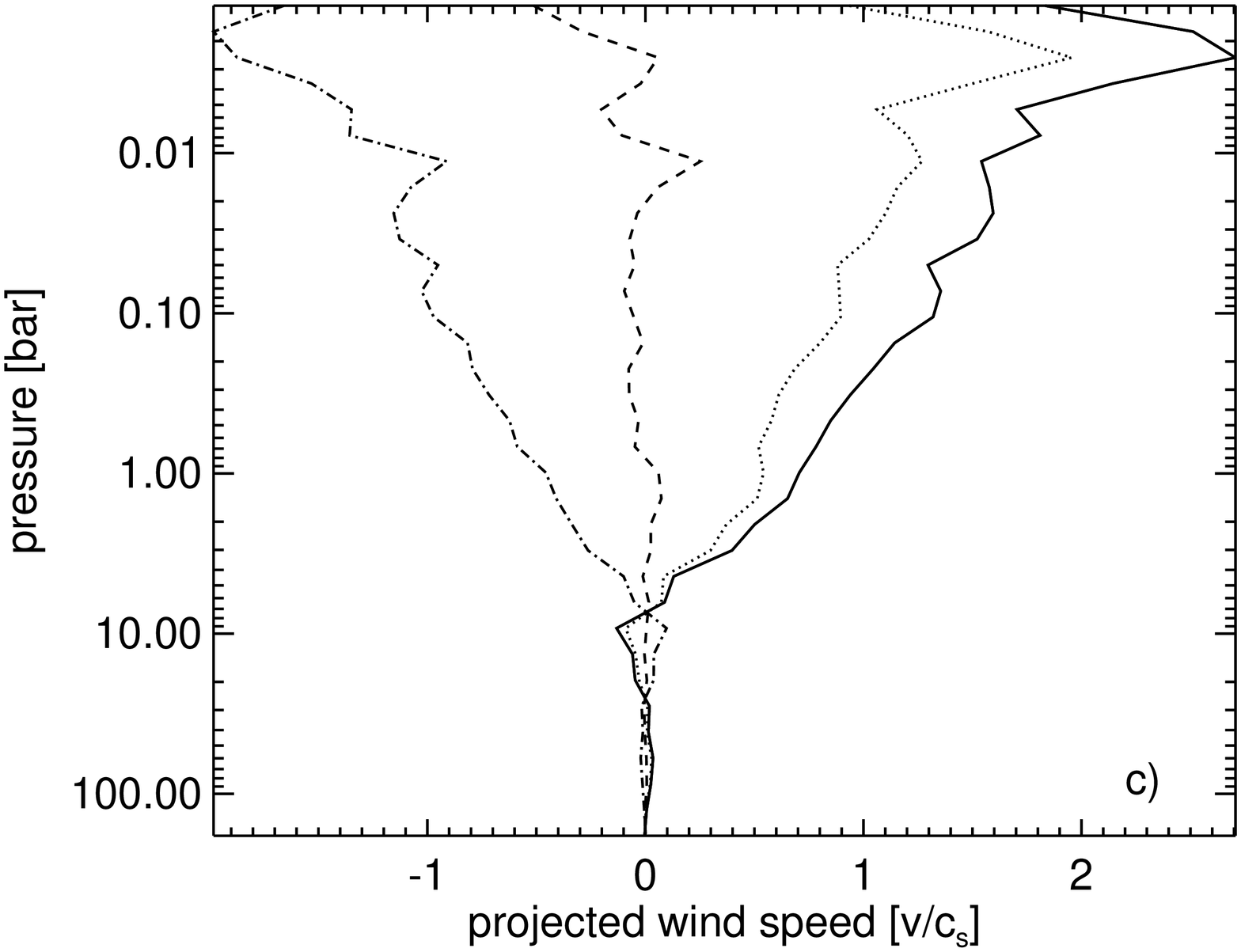}
\includegraphics[angle=90,scale=0.32]{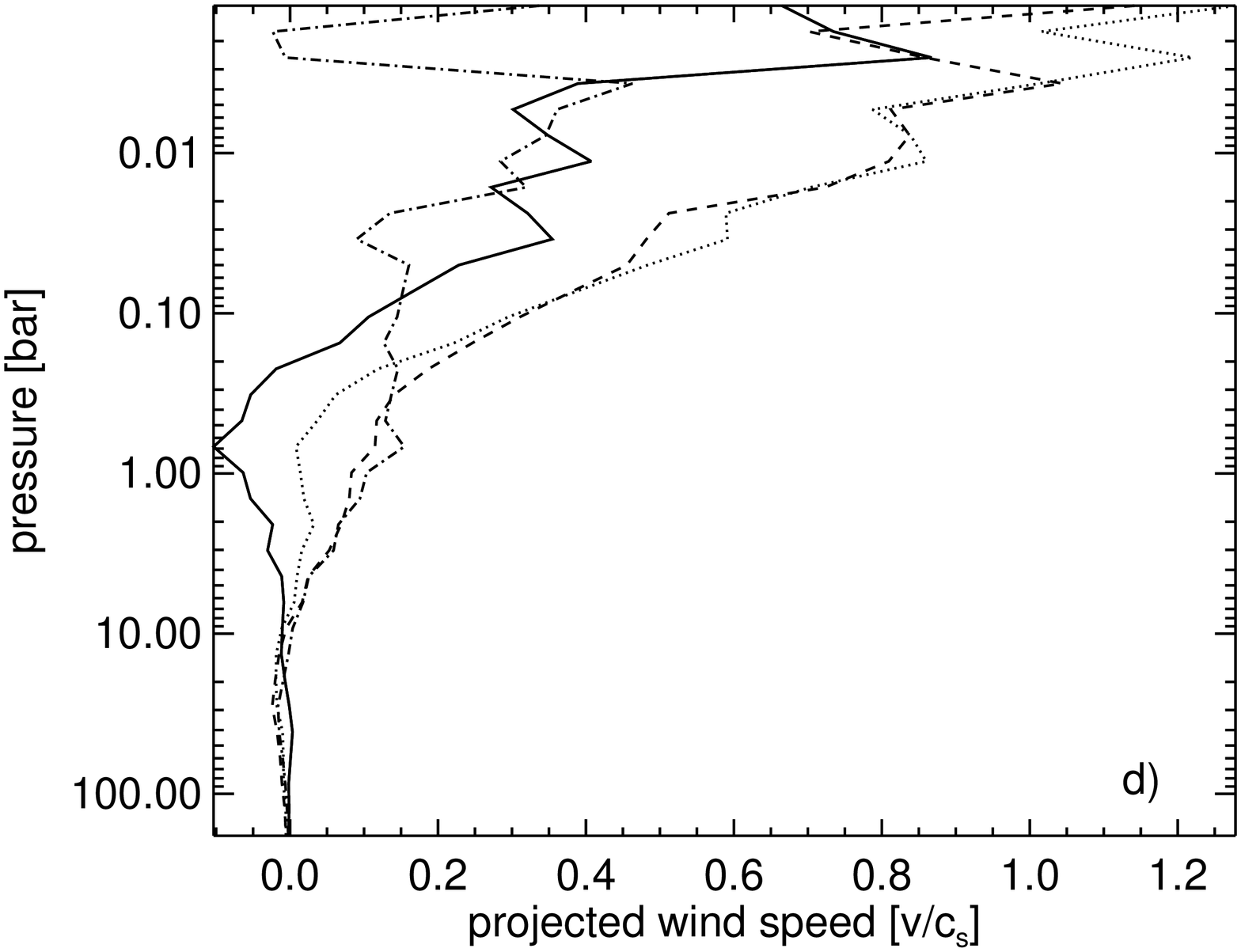}
\caption{Representative profiles of wind velocities, in units of the
local sound speed $c_s$, projected along various optical paths at a
fixed zenith angle $\theta_0=53$ degrees in the atmospheric
circulation model of \citet{RauMen09}. The various panels show
profiles at the model substellar point (a), antistellar point (b),
west equatorial terminator (c) and north pole (d). In each panel, the
various curves show projected velocity profiles for optical paths
oriented to the east (solid line), north-east (dotted), north (dashed)
and north-west (dash-dotted). Velocity differentials which are a
sizable fraction of, and in some cases exceed, $c_s$ are typical.}
\label{fig:one}
\end{figure}

Figure~\ref{fig:one} shows profiles of wind velocity projected along
representative optical paths, as a function of pressure, $p$ (a proxy
for height in the atmosphere). All velocities are expressed in units
of the local value of the adiabatic sound speed, $V_{\rm proj}(p)/
c_s(p;T) $. The same $H_2$-dominated atmospheric gas parameters as in
\citet{RauMen09} are used to calculate the sound speed: $c_s \equiv
\sqrt{\gamma {\cal R} T}$, with an adiabatic index $\gamma =
1/(1-\kappa)$, $\kappa=0.321$ and a gas constant ${\cal R}=4593$ J
kg$^{-1}$ K$^{-1}$. The various panels show profiles at the model
substellar point (a), antistellar point (b), west equatorial
terminator (c) and north pole (d). In each panel, the various curves
show projected velocity profiles for optical paths oriented to the
east (solid line), north-east (dotted), north (dashed) and north-west
(dash-dotted) on the sphere. Projected velocity profiles for other
cardinal directions can be deduced by symmetry.

Figure~\ref{fig:one} reveals significant gradients in projected wind
velocity as one crosses the atmosphere along representative slanted
optical paths for thermal radiation. Velocity differentials over one
pressure scale height easily amount to $\sim 0.2$- $0.5~c_s$ and they
exceed $c_s$ in some cases, especially high up in the
atmosphere. Velocity differentials $\ga c_s$ are typical when crossing
several pressure scale heights.

The different panels in Figure~\ref{fig:one} illustrate the diverse
character of projected velocity profiles at various locations around
the planet. Furthermore, panels a), b) and c) exemplify the
anisotropic nature of these projected velocity profiles. Projected
velocity gradients are systematically weak in the north (or
equivalently south) direction but they can become significant when the
east-west direction is sufficiently sampled by the optical path under
consideration.  This is easily understood as resulting from the
predominantly zonal (east-west) nature of winds in this and other hot
Jupiter atmospheric circulation models \citep{RauMen09,SMC08}. It is
then clear that the azimuthal angular phase space for thermal
radiation will be dominated by optical paths sampling significant
velocity gradients, with unusually low gradient values relevant only
for the small fraction of all paths that are closely aligned with the
north-south direction. As discussed in \S\ref{S:dynatm}, to the extent
that projected velocity gradients of the magnitude shown in
Fig.~\ref{fig:one} impact the transport of thermal radiation, the
radiative transfer problem will become anisotropic via the sampling of
a variety of azimuthal and zenith angles, even when the Planck source
function itself is isotropic (in the fluid frame).

\subsection{Line Shapes and Widths} \label{sec:widths}

Opacity sources in hot Jupiter atmospheres are largely dominated by
discrete atomic and molecular lines, particularly in the thermal
portion of the atmospheric radiation spectrum
\citep[e.g.,][]{SharpBurrows07,Freedman08}. Lines are broadened well
beyond their quantum mechanical width through Doppler shifts
associated with the thermal motions of atomic and molecular
constituents (``Doppler broadening'') and through the effect of
collisions of these atoms and molecules with other gas constituents
(``pressure broadening''). The relative width of Doppler and pressure
broadening depends on the local conditions of density and temperature
in the atmosphere, as well as on the specific radiative constituent
under consideration. The general shape of a line that is both Doppler-
and pressure-broadened, i.e. the detailed functional form of the line
function $\Phi$ in eq.~(\ref{eq:one}), is given by the Voigt function
\citep{GoodyYung89,ThomasStamnes02,Liou02}.

Intuitively, one expects the radiation transport in an atmosphere with
strongly pressure-broadened lines and very subsonic wind speeds to be
relatively insensitive to the additional Doppler shifts due to bulk
flow velocities projected along various optical paths.  Indeed, the
typical thermal width of a Doppler-broadened line is closely related
to the sound speed of the atmospheric gas (see eq.~[\ref{eq:DnuD}]
below), so that bulk line shifts should only minimally deform line
shapes in the limit of very subsonic wind speeds. This, combined with
the dominance of pressure broadening, is the main justification behind
the use of a static treatment for radiation transport in Solar System
planetary atmospheres \citep[e.g.,][]{GoodyYung89}. To evaluate the
possible role of bulk Doppler shifts for radiation transport in hot
Jupiter atmospheres, it is thus important to evaluate the shapes and
widths of radiative lines in these atmospheres.

Various static, globally-averaged radiative transfer studies have
established that the radiatively forced regions of hot Jupiter
atmospheres are found, broadly speaking, above the 10--100 bar
pressure level
\citep[e.g.,][]{Seager1998,Sudarsky2000,Barman2005,Iro05,Seager2005}. These
models also indicate that photospheric levels\footnote{The photosphere
can be defined as the height in the atmosphere at which photon escape
to space becomes possible, i.e., where the monochromatic optical
thickness approaches unity.} vary strongly with wavelength in the
optical--IR spectral range, as a result of strong variations in line
opacities, from a few bars to $\la 10^{-2}$~bars typically
\citep[e.g.,][]{Seager2005,SharpBurrows07}. Regions in the pressure
range from $\sim 10$ to $10^{-3}$ bars are thus of particular interest
for the study of radiation transport in dynamic hot Jupiter
atmospheres.

A detailed account of the typical treatment of lines in static
radiative transfer calculations for hot Jupiter atmospheres is
provided by \citet{SharpBurrows07}. The general Voigt profile of a
pressure- and Doppler-broadened line is given by

\begin{equation} \label{eq:voigt}
\Phi_V (\nu - \nu_0) = \frac{1}{\pi^{3/2}} \frac{\Delta \nu_p}{\Delta \nu_D} \int_{-\infty}^{+\infty} \frac{1}{(\nu'-\nu_0)^2 + \Delta \nu_p^2} \exp \left[-\frac{(\nu-\nu')^2}{\Delta \nu_D^2}\right] d\nu',
\end{equation}
where $\Delta \nu_p$ and $\Delta \nu_D$ measure the pressure and
Doppler broadening widths, respectively, and $\Phi_V$ is normalized to
unity \citep{GoodyYung89,ThomasStamnes02,Liou02}. Pressure broadening
generally depends on pressure, temperature and the radiative
constituent under consideration but the ``classical'' scaling

\begin{equation}
\Delta \nu_p = 0.02 - 0.05 \left( \frac{P}{{\rm 1~bar}} \right) \left( \frac{T}{{\rm 1500~K}}\right)^{-1/2}~{\rm cm}^{-1},
\end{equation}
with a $-1/2$ power law dependence on temperature, should be
sufficient for our order-of-magnitude estimates
\citep{GoodyYung89,ThomasStamnes02,Liou02,SharpBurrows07}. Doppler
broadening depends on temperature, wavelength ($\lambda_0 = c /
\nu_0$) and the mass, $m_{\rm mol}$, of the radiative constituent
under consideration,

\begin{equation}  \label{eq:DnuD}
\Delta \nu_D \equiv \frac{\nu_0}{c} \sqrt{\frac{2 k T}{m_{\rm mol}}} \simeq \frac{\nu_0}{c} \sqrt{\frac{2 m_{\rm H_2}}{\gamma m_{\rm mol}}} c_s \simeq 0.14 \left( \frac{m_{\rm H_2}}{m_{\rm mol}} \right)^{1/2} \left( \frac{T}{{\rm 1500~K}}\right)^{1/2}  \left( \frac{\lambda_0}{{\rm 1~\mu m}}\right)^{-1} ~{\rm cm}^{-1},
\end{equation}
where $m_{\rm H_2}$ is the mass of the ${\rm H_2}$ molecule (the
dominant atmospheric constituent), $\gamma$ is the gas adiabatic index
and $c_s$ is the corresponding adiabatic sound speed.  This scaling
illustrates how the Doppler width is reduced for radiative
constituents which are typically more massive than molecular
hydrogen. Note that the linear scaling of $\Delta \nu_D$ with the
central wavelength $\lambda_0$ implies significant Doppler width
variations across the relevant optical-IR spectral range.

Substantial variations in temperature on constant pressure levels are
found, from day- to night-side, in current atmospheric circulation
models for hot Jupiters
\citep{CS05,CS06,Langton2007,DobbsDixon2008,S08,S09,MR08,RauMen09}. Here,
for simplicity, we choose $T=1500$~K and $1000$~K as representative
temperature values at the $1$ bar and $10^{-2}$ bar levels,
respectively \citep[see, e.g., Fig.~3 of][]{RauMen09}. The ratio of
Doppler to pressure broadening widths is given by

\begin{equation}
\frac{\Delta \nu_D}{\Delta \nu_p}= 3-7 \left( \frac{m_{\rm H_2}}{m_{\rm mol}} \right)^{1/2} \left( \frac{T}{{\rm 1500~K}}\right)  \left( \frac{P}{{\rm 1~bar}} \right)^{-1} \left( \frac{\lambda_0}{{\rm 1~\mu m}}\right)^{-1}.
\end{equation}

For relevant molecules, typical values of the mass ratio factor,
$(m_{\rm H_2}/m_{\rm mol})^{1/2}$, are $\simeq 1/2.8{\rm ~(CH_4)} $,
$1/3~{\rm (H_2O)}$, $1/4.7{\rm ~(CO_2)}$ and $1/5.6{\rm ~(TiO)}$.  At
1 bar, for $T=1500$~K, the typical broadening ratio is thus $\Delta
\nu_D /\Delta \nu_p \simeq 0.5-2$ at ${\rm 1~\mu m}$ and $0.05-0.2$ at
${\rm 10~\mu m}$.  At $10^{-2}$ bar and $T=1000$~K, the ratio becomes
$\Delta \nu_D /\Delta \nu_p \simeq 30-130$ at ${\rm 1~\mu m}$ and
$3-13$ at ${\rm 10~\mu m}$. Doppler broadening is thus significant at
the 1 bar level, especially in the near-IR and the optical, and it
becomes increasingly dominant across the entire optical-IR spectral
range higher up in the atmosphere.

The sizable contribution of thermal Doppler broadening to the width of
radiative lines at photospheric levels in hot Jupiter atmospheres,
together with the scaling $\Delta \nu_D < c_s \times \nu_0 / c$ from
Eq.~(\ref{eq:DnuD}), suggests that bulk Doppler shifts from
atmospheric motions near the sound speed could modify the shapes of
radiative lines significantly. By contrast, much deeper in the
atmosphere, where $\Delta \nu_D / \Delta \nu_p \ll 1$, Doppler shifts
from bulk atmospheric motions near the sound speed would only amount
to small shifts over comparatively wide, pressure-broadened
lines. Furthermore, wind speeds themselves may be reduced at these
deeper levels (see, e.g., Fig.\ref{fig:one}).

These qualitative arguments are not very informative about the
possible consequences of bulk Doppler shifts on radiation transport in
a dynamic atmosphere. In particular, since the Doppler cores of
radiative lines are often very optically thick (``saturated'') in hot
Jupiter and other planetary atmospheres, they do not necessarily
contribute much to the overall atmospheric energy budget, at least in
the static case.\footnote{Although this is not always explicitly
stated, line-by-line radiative transfer models for hot Jupiter
atmospheres currently available in the literature, with typical
spectral resolutions $\sim 1$~cm$^{-1}$
\citep[e.g.,][]{SharpBurrows07,Seager1998,Seager2005}, do not
necessarily resolve the narrow Doppler cores of radiative lines, with
typical widths $\Delta \nu_D \ll 1$~cm$^{-1}$ according to
Eq.~(\ref{eq:DnuD}).  } To help us evaluate more quantitatively the
magnitude of bulk Doppler shift effects, we now turn to models of line
transmissivities and equivalent widths in dynamic atmospheres.

\section{Transmissivity and Equivalent Width Models} \label{S:EW}

\subsection{Transmissivities}

As summarized in \S\ref{S:dynatm}, monochromatic transmissivities
encapsulate the absorption and emission properties of the atmospheric
medium (eqs.~[\ref{eq:two}--\ref{eq:three}]). Here, we isolate the
effects of velocity gradients along an arbitrary optical path by
modeling the monochromatic transmissivity in a dynamic atmosphere as

\begin{equation} \label{eq:elev}
{\cal T}_\nu= \exp \left[ - \int_0^1 \tilde{k}_{\rm tot}
\Phi_V \left[ (\nu - \nu_0) + \frac{\nu_0}{c} V_{\rm proj} \times s')
\right] ds'\right],
\end{equation}
where $s'$ is the length along the unit optical path,\footnote{In
  Eq.~(\ref{eq:elev}), $\tilde{k}_{\rm tot}$ has unit of inverse
  length. Without loss of generality, we only consider optical paths
  of unit length in our simplified model. As a result, $\tilde{k}_{\rm
    tot}$ fully characterizes the optical thickness of the modeled
  path. To differentiate the more general line integrated opacity
  ${k}_{\rm tot}$ appearing in Eq~(\ref{eq:four}), which has units of
  cross-section per unit mass, from the simpler formulation adopted
  above, we write it as $\tilde{k}_{\rm tot}$ in the simplified
  model.} the line integrated opacity $\tilde{k}_{\rm tot}$ is assumed
to be constant along the path and $\Phi_V$ is the dimensionless Voigt
line profile defined in Eq.~(\ref{eq:voigt}).  The above expression
for the bulk Doppler shift term in the line function assumes a
constant velocity gradient along the path, i.e. a velocity offset that
increases linearly with $s'$, from $0$ at $s'=0$ to the maximum value
$V_{\rm proj}$ at $s'=1$. This simple model isolates the effects of
bulk Doppler shifts by assuming that the path is otherwise
homogeneous. In a more realistic atmospheric model, optical paths
would be inhomogeneous, with line strengths, $\tilde{k}_{\rm tot}$,
generally varying with temperature and line shapes, $\Phi_V $,
generally varying with both temperature and pressure along the
specific path \citep{GoodyYung89,ThomasStamnes02,Liou02}.\footnote{In
  principle, the combination of Eqs.~(\ref{eq:three}),~(\ref{eq:four})
  and~(\ref{eq:voigt}) requires one to account for the profile
  variation with pressure and temperature separately for each
  individual line along the optical path of interest. A very common
  simplifying assumption in the treatment of atmospheric radiation,
  known as the Curtis-Godson approximation, is to treat the optical
  path as if it were homogeneous, i.e. with constant pressure and
  temperature, and use adequately path-averaged values for the line
  strength, shape and the absorber amount
  \citep[e.g.,][]{RodgWal66,GoodyYung89,ThomasStamnes02,Liou02}. While
  the accuracy of the Curtis-Godson approximation has been extensively
  tested in the context of static atmospheric radiation transport, its
  possible breakdown when optical paths acquire anisotropic properties
  in the presence of bulk Doppler shifts could lead to subtle
  complications in the treatment of radiation transport in a dynamic
  atmosphere. This specific aspect of the problem is not explicitly
  addressed by our simple transmissivity models.}

\begin{figure}
\begin{center}
\plotone{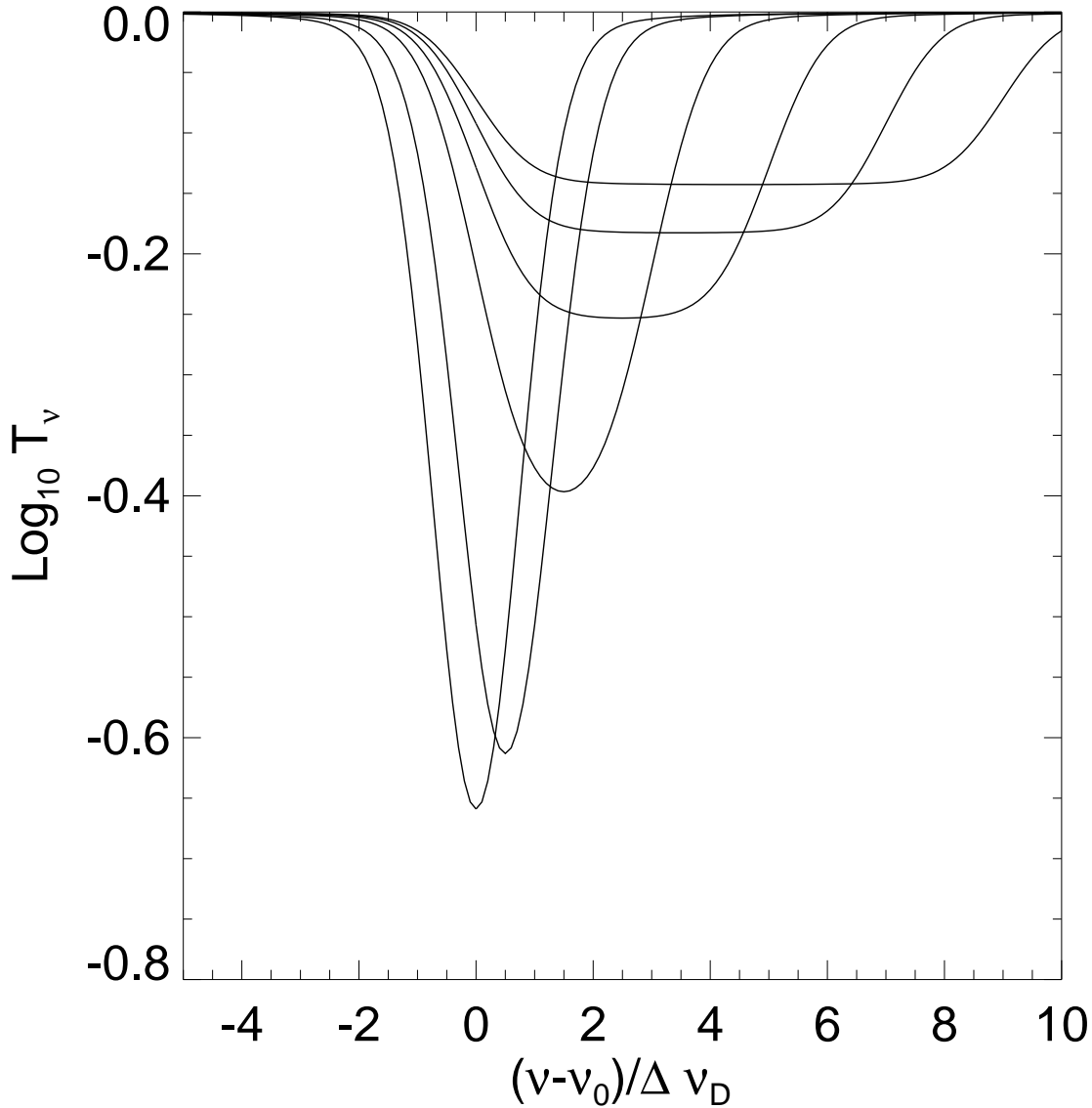}
\caption{Monochromatic transmissivity curves for a line centered at
$\nu=\nu_0$ with a Voigt profile shape and a fixed ratio of Doppler-
to pressure-broadening widths, $\Delta \nu_D / \Delta \nu_p =10$. From
left to right, transmissivity profiles are shown for linear bulk
velocity gradients with total offsets $V_{\rm proj} / \Delta \nu_D =
0$ (static), $1,3,5,7$ and $9$~$ \times c / \nu_0$ over the full
optical path. The specific model shown here is marginally
optically-thick, with a line-integrated absorption coefficient $\tilde{k}_{\rm
tot} =3$ over the full optical path, but these transmissivity profiles
are representative of more general cases. Monochromatic
transmissivities are significantly affected by bulk Doppler shifts in
excess of a few thermal Doppler widths. The increasingly blue-shifted
transmissivity profiles shown have equivalent widths exceeding that of
the static profile by $2.4, 14, 24, 31$ and $35 \%$, from left to
right.}
\label{fig:two}
\end{center}
\end{figure}

Figure~\ref{fig:two} shows representative profiles of monochromatic
transmissivities for a single line centered at $\nu=\nu_0$ in the
model described by Eq.~(\ref{eq:elev}). Fixed values for the line
integrated opacity, $\tilde{k}_{\rm tot}=3$, and for the ratio of Doppler to
pressure broadening widths, $\Delta \nu_D / \Delta \nu_p =10$, were
adopted.  The deepest transmissivity curve corresponds to the static
reference model, with $V_{\rm proj}=0$. In decreasing order of profile
depth, from left to right, the other curves correspond to models with
total velocity offsets $V_{\rm proj} =1,3,5,7$ and $9$ $\Delta \nu_D
\times c / \nu_0$, i.e., in units of the thermal Doppler width.

From the point of view of absorption, transmissivities ${\cal T}_\nu
\la 1$ correspond to the minimal absorption of the optically thin
regime, while ${\cal T}_\nu \ll 1$ correspond to the strong absorption
limit of the optically thick regime. As is clear from
Figure~\ref{fig:two}, monochromatic transmissivities can be
considerably affected by bulk Doppler shifts approaching or exceeding
the thermal Doppler width, $\Delta \nu_D$, of the line under
consideration. A switch from partially optically-thick to fully
optically thin occurs in Figure~\ref{fig:two}. While only the case
with $\tilde{k}_{\rm tot}=3$ is shown, the shift and the flattening of line
transmissivities seen here is qualitatively representative of what is
seen for more general cases, with values of $\tilde{k}_{\rm tot}$ ranging from
$ \ll 1$ to $\gg 1$. We note that the flat-top nature of line
transmissivities shown in Figure~\ref{fig:two} when velocity gradients
are large, i.e., $V_{\rm proj} \gg \Delta \nu_D \times c / \nu_0$, is
consistent with the flattening discussed by \citet{Castor70} using the
Sobolev approximation (see his Fig.~4).

It is worth emphasizing that the magnitude of the velocity gradients
considered in Figure~\ref{fig:two} are relevant to hot Jupiter
atmospheres. Writing the bulk Doppler shift term in
Equation~(\ref{eq:one}) as $\Delta \nu_{\rm bulk}= \bb{n} \bcdot
\bb{v(r)} \times \nu_0 / c$, we can express its overall magnitude in
our simple transmissivity model in terms of the thermal Doppler width,
$\Delta \nu_D$, as

\begin{equation}
\frac{\Delta \nu_{\rm bulk}}{\Delta \nu_D} = \sqrt{\frac{\gamma m_{\rm mol}}{2 m_{\rm H_2}}} \frac{V_{\rm proj}}{c_s}.
\end{equation}

Since, as seen earlier, the square root factor can reach values up to
5--6 for relevant molecules and, as suggested by Figure~\ref{fig:one},
differentials of projected velocities can reach up to $V_{\rm proj}
\sim$ 1--2 $c_s$ along some atmospheric optical paths, the entire
range of velocity offset values considered in Figure~\ref{fig:two} is
probably of interest for hot Jupiter atmospheres.

By themselves, offsets and distortions of line transmissivities of the
type shown in Figure~\ref{fig:two} are not observationally too
significant because they occur over spectral intervals typically $<
1$~cm$^{-1}$, which is well below the effective resolving power
achieved in optical-IR spectroscopic studies of hot Jupiter
atmospheres. To the extent that such bulk Doppler distortions can
alter the radiative energy balance in the dynamic atmosphere, however,
they could in principle modify the radiative structure of hot Jupiter
atmospheres. To evaluate the magnitude of this effect, we turn to a
discussion of line equivalent widths.

\subsection{Equivalent Widths} \label{sec:EW}

Let us first justify our use of equivalent widths by borrowing from
the existing literature on ``narrow-band'' spectral models
\citep{GoodyYung89,ThomasStamnes02,Liou02}. Ignoring the boundary term
for simplicity, the integral term for the intensity given in
Eq.~(\ref{eq:two}) is essentially of the form

\begin{equation}
I_\nu = \int_{0}^{1} B_\nu d {\cal T}_\nu,
\end{equation}
which is simply reformulating it as the cumulative Planck emission and
its absorption along a specific optical path, weighed by the
transmissivities of the various contributing layers, from closely
adjacent ones (with transmissivity ${\cal T}_\nu \simeq 1$) to more
distant ones (with ${\cal T}_\nu \simeq 0$, in the optically thick
limit). In narrow-band models, the frequency-integrated intensity is
written

\begin{equation} \label{eq:narrowb}
I = \int_{0}^{\infty} d\nu \int_{0}^{1} B_\nu d {\cal T}_\nu \simeq \sum_{i} \Delta \nu_i  \int_{0}^{1} B_i d \bar{{\cal T}}_i,
\end{equation}
where $\Delta \nu_i$ is the frequency span of the $i$th narrow band,
the Planck function value $B_i$ is considered to be a constant over
each narrow band and the band-averaged transmissivities are defined by

\begin{equation} \label{eq:narrowb2}
\bar{{\cal T}}_i  \equiv  \frac{1}{\Delta \nu_i}  \int_{\Delta \nu_i} {\cal T}_\nu d\nu.
\end{equation}

The narrow-band formalism is well justified as long as the number of
narrow bands is sufficiently large for the Planck function to be well
approximated by a constant in each band
\citep{GoodyYung89,ThomasStamnes02,Liou02}. Since atmospheric fluxes
are obtained by angular integration of the intensity,
equation~(\ref{eq:narrowb}) relates the radiative energy balance of
the atmosphere to multiple Planck-weighted integrals of band-averaged
transmissivities.

The narrow-band formalism is traditionally used with bands that are
still wide enough to encompass a large number of individual radiative
lines. By contrast, we use it here in its simplest formulation, with
very narrow bands including only a single spectral line, assuming that
each such line is well isolated from all the other lines \citep[see,
e.g.,][chapter 4]{GoodyYung89}. In that limit, the equivalent width
(EW) of a specific line can be written

\begin{equation} \label{eq:ew}
EW= \int_{\Delta \nu_i} (1- {\cal T}_\nu) d\nu = \Delta \nu_i (1 -\bar{{\cal T}}_i),
\end{equation}
where $\Delta \nu_i$ now represents the typical spacing between
neighboring isolated lines. It should be noted that the $EW$ is
independent of the value of $\Delta \nu_i$ adopted as long as the
spectral interval is large enough to encompass all of the
transmissivity values contributing meaningfully to the above integral
(i.e., ${\cal T}_\nu < 1$). Thus, when line overlap can be omitted,
Eqs.~(\ref{eq:narrowb}--\ref{eq:ew}) clarify the direct relation that
exists between the radiative energy balance of an atmosphere ($\propto
I$), Planck-weighted integrals of band-averaged transmissivities
($\propto d\bar{{\cal T}}_i$) and the equivalent widths of all
important radiative lines ($d\bar{{\cal T}}_i \propto -dEW $).  This
is our main justification for using deviations in line equivalent
widths as a quantitative measure of the effects of bulk Doppler shifts
on the overall radiative balance of a dynamic atmosphere.

We perform equivalent width calculations by integrating transmissivity
profiles like the ones shown in Figure~\ref{fig:one}, following
equation~(\ref{eq:ew}). For the calculations presented here, we
verified that our EW results are independent of the spectral interval
chosen for integration, as long as it spans more than 100--1000
thermal Doppler widths ($\Delta \nu_D$) on each side of the line
central frequency, $\nu_0$. We evaluate the effects of bulk Doppler
shifts simply by comparing the equivalent widths of lines in a static
atmosphere, $EW_0= EW(V_{\rm proj}=0)$, to those obtained in the
presence of finite bulk Doppler shifts ($V_{\rm proj} \neq 0$).

In all our calculations, using the simple transmissivity model
described by Eq.~(\ref{eq:elev}), we find that EWs in dynamic
atmospheres are systematically larger than the corresponding value in
a static atmosphere, $EW_0$. We quantify this trend with the
fractional excess, $EW/EW_0-1$, expressed in percents. For example,
for the various shifted transmissivity profiles shown in
Figure~\ref{fig:two}, we find EW excesses of $2.4, 14, 24, 31$ and $35
\%$ over the static $EW_0$ value for the profiles with total velocity
offsets $V_{\rm proj} =1,3,5,7$ and $9$~$\Delta \nu_D \times c /
\nu_0$, respectively.

\begin{figure}
\begin{center}
\plotone{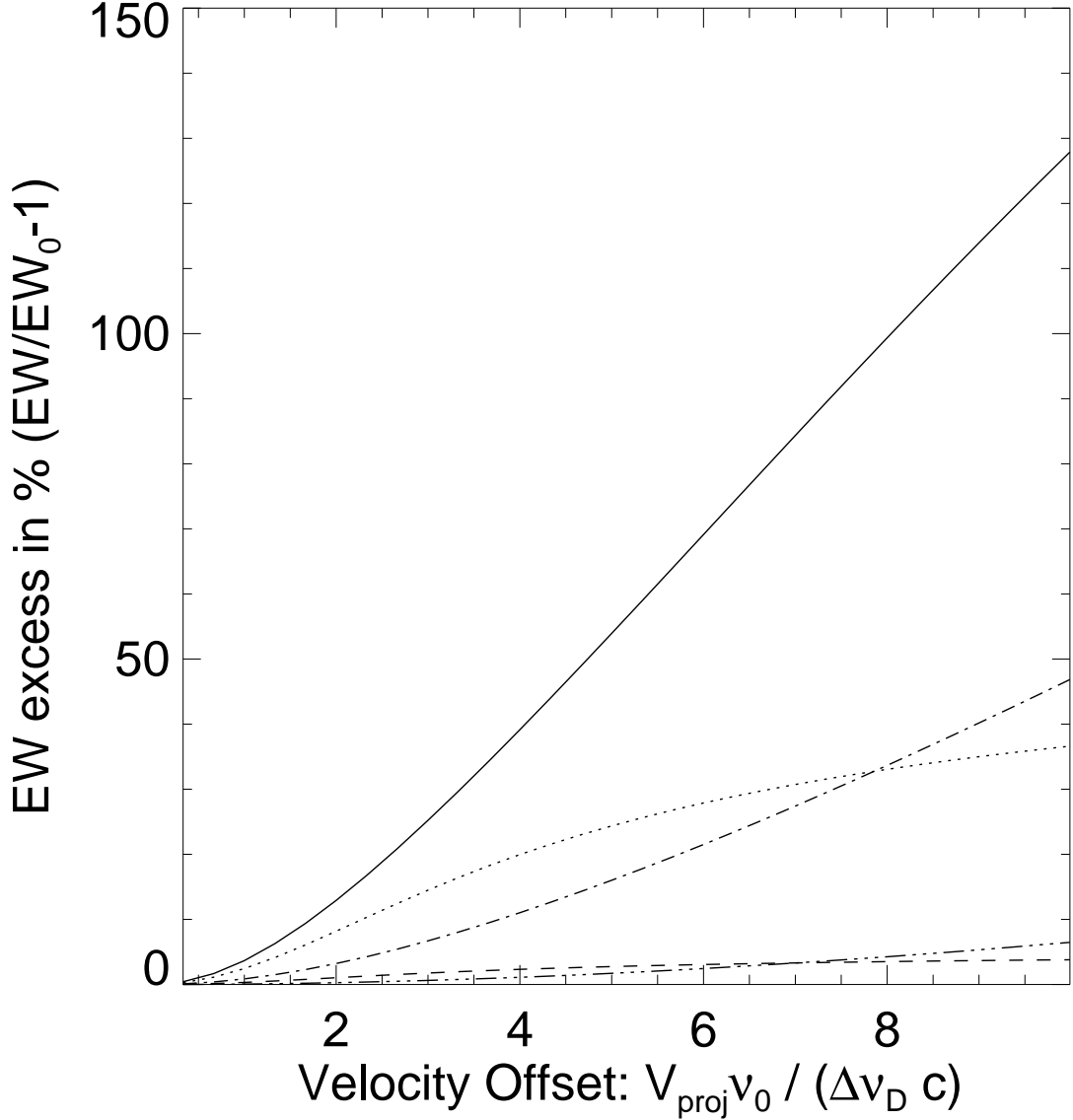}
\caption{Equivalent width excess (in \%) of a radiative line in a
dynamic atmosphere with a linear bulk velocity gradient, as a function
of $V_{\rm proj}$, the total velocity offset over the optical path, in
units of the line's thermal Doppler width ($\Delta \nu_D \times
c/\nu_0$). In this specific model, the line has a Voigt profile shape
with a fixed ratio of Doppler- to pressure-broadening widths, $\Delta
\nu_D / \Delta \nu_p =10$.  The various curves show results for lines
with different total absorption coefficients: $\tilde{k}_{\rm tot}= 0.3$
(dashed line), $3$ (dotted), $30$ (solid), $300$ (dash-dotted) and
$3000$ (triple-dot dashed). }
\label{fig:three}
\end{center}
\end{figure}

Figure~\ref{fig:three} further clarifies the variation of the EW
excess with the magnitude of the total velocity offset, $V_{\rm
proj}$, in units of $\Delta \nu_D \times c / \nu_0$, in a model with
fixed ratio of broadening widths, $\Delta \nu_D/\Delta \nu_p =10$, for
the Voigt line function. The various curves show models with line
integrated absorption coefficient $\tilde{k}_{\rm tot}= 0.3$ (dashed line),
$3$ (dotted; same as in Fig.~\ref{fig:two}), $30$ (solid), $300$
(dash-dotted) and $3000$ (triple-dot dashed). While the EW excesses
increase with the value of $V_{\rm proj}$, the exact dependence is not
entirely trivial. For $\tilde{k}_{\rm tot}= 3$, $30$ and $300$, EW excesses
from several tens of percents to more than a hundred percent are
possible, for the same range of $V_{\rm proj}$ values as deemed
relevant for hot Jupiter atmospheres earlier.

\begin{figure}
\begin{center}
\plotone{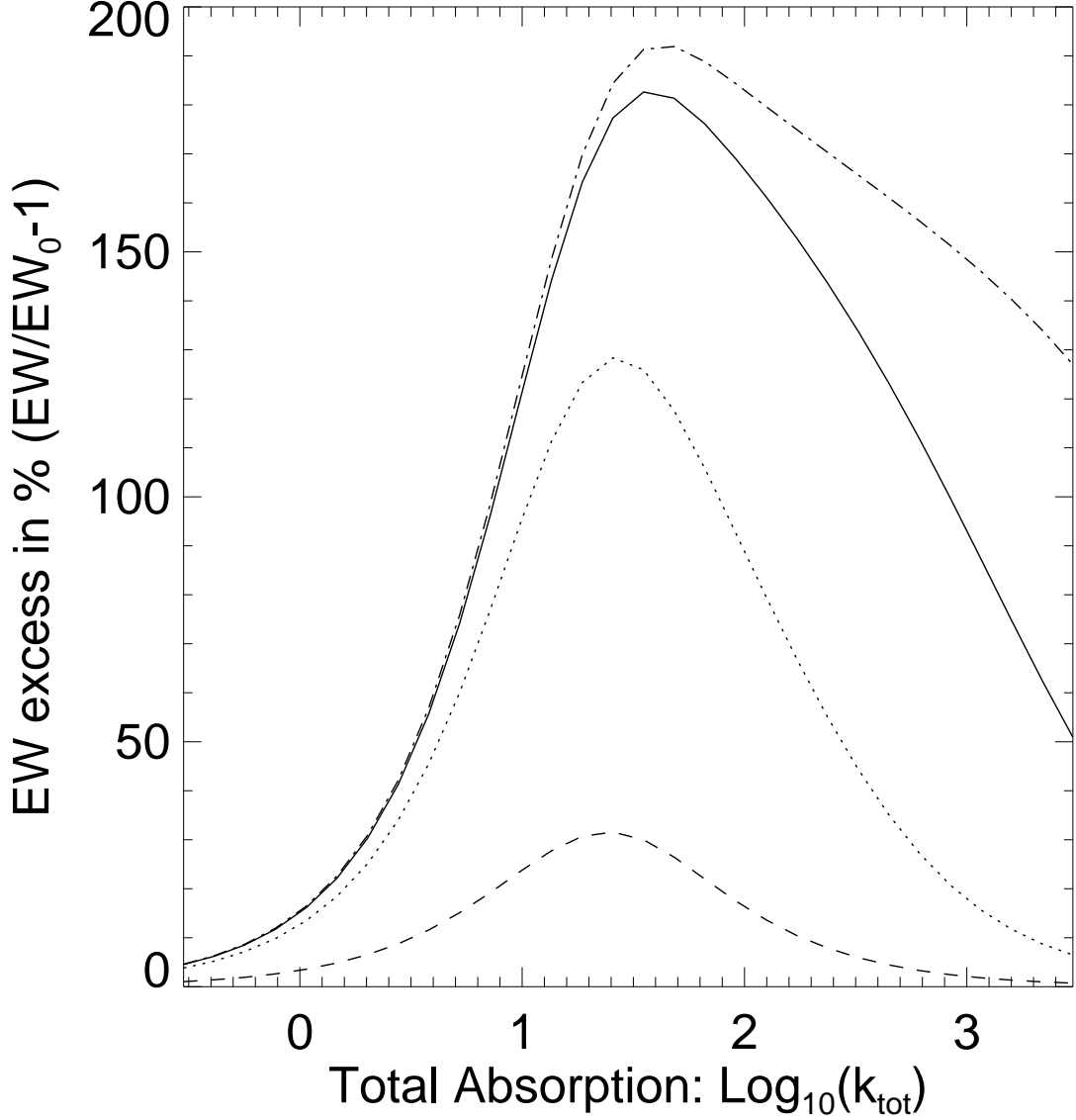}
\caption{Equivalent width excess (in \%) of a radiative line in a
dynamic atmosphere with a linear bulk velocity gradient, as a function
of the total line absorption coefficient, $\tilde{k}_{\rm tot}$. In this
specific model, a relatively large value of the total velocity offset
over the full optical path, $V_{\rm proj} =10~\Delta \nu_D \times
c/\nu_0$, is adopted. The line has a Voigt profile shape and the
various curves shown correspond to different ratios of the Doppler- to
pressure-broadening widths: $\Delta \nu_D / \Delta \nu_p =1$ (dashed
line), $10$ (dotted), $100$ (solid) and $1000$ (dash-dotted). }
\label{fig:four}
\end{center}
\end{figure}

Figure~\ref{fig:four} shows the systematic variation of the EW excess
with the magnitude of the total line absorption coefficient, $\tilde{k}_{\rm
tot}$, in a model with a value of the total velocity offset fixed at
$V_{\rm proj} =10$~$\Delta \nu_D \times c / \nu_0$. The various curves
show models with Voigt function broadening ratios $\Delta \nu_D/\Delta
\nu_p =1$ (dashed line), $10$ (dotted; same as in Figs.~\ref{fig:two}
and~\ref{fig:three}), $100$ (solid) and $1000$ (dash-dotted). The EW
excess first increases with the value of $\tilde{k}_{\rm tot}$, peaks around
$\tilde{k}_{\rm tot} \sim 20$-$40$ and then drops at larger $\tilde{k}_{\rm tot}$
values. Peak EW excesses are thus reached for rather strong optically
thick conditions. Note that EW excesses $\sim 150$-$200\%$ are
possible for $\tilde{k}_{\rm tot} \sim 30$, $\Delta \nu_D/\Delta \nu_p \ga
100$ and $V_{\rm proj} \sim 10$~$\Delta \nu_D \times c / \nu_0$. EW
excesses easily reach several tens of percents at $\tilde{k}_{\rm tot} \sim
30$ in all of the models considered here, for the rather large value
of $V_{\rm proj} =10$~$\Delta \nu_D \times c / \nu_0$ adopted (see
again Fig.~\ref{fig:three} for the dependence with $V_{\rm proj}$).

We can understand various trends in the behavior of the EW excesses
with $ V_{\rm proj}$, $\tilde{k}_{\rm tot}$ and $\Delta \nu_D/\Delta
\nu_p$ in our simple models as follows. As already mentioned earlier,
we have found that the shifts and distortions of monochromatic
transmissivity profiles shown in Fig.~\ref{fig:two} are qualitatively
representative of the general behavior seen in all our models, whether
the line is in the optically thin or optically thick regime. For a
line well into the optically thick regime ($1- {\cal T}_{\nu_0} \simeq
1$), the growth in EW with $ V_{\rm proj}$ can be simply understood as
a stretching of the optically thick portion of the transmissivity
curve, which contributes maximally to the growth in EW according to
the integral in equation~(\ref{eq:ew}). This is the main trend
observed for moderate to high values of $\tilde{k}_{\rm tot}$ in
Figure~\ref{fig:three}.

In the optically thin regime, the increase in EW excess with
$\tilde{k}_{\rm tot}$ can be understood as resulting from the presence
of larger radiative intensities outside the line center (where maximum
absorption occurs), so that Doppler-shifted absorption around the line
center contributes increasingly to the EW excess. As the value of
$\tilde{k}_{\rm tot}$ is increased well into the optically thick
regime, however, the much shallower pressure-broadened wings of the
Voigt line profile start making a significant contribution to the EW
integral, over an increasingly wider range of frequencies. This
reduces the effective contribution of the optically-thick, shifted
Doppler core to the total line EW, which explains the decline in EW
excess at large $\tilde{k}_{\rm tot}$ values in Figure~\ref{fig:four},
the more so in models with larger values of the pressure-broadening
parameter, $\Delta \nu_p$.

\section{Discussion and Conclusion} \label{S:impl}

Our main result has been to establish the possibility that bulk
Doppler shifts from fast, sonic or transonic wind speeds in hot
Jupiter atmospheres could affect radiation transport and thus in
principle influence the overall radiative structure of these
atmospheres. We have quantified the magnitude of these effects with
simplified transmissivity models in a dynamic atmosphere, by assuming
that linear bulk velocity gradients are present along otherwise
homogeneous optical paths and by adopting Voigt line profile
parameters appropriate for the radiatively forced regions of hot
Jupiter atmospheres. We have found that bulk Doppler shifts
systematically increase the equivalent widths of isolated radiative
lines, relative to the equivalent widths in a static atmosphere, with
excesses easily reaching several tens of percents for relevant model
parameters.

Flux errors below the few percent level may well be acceptable in
atmospheric models currently used to interpret the increasingly rich
set of observational data on hot Jupiter atmospheres since other
important atmospheric unknowns also contribute to modeling errors
(e.g., the exact compositional profiles of radiatively active
constituents). Flux errors at the level of several tens of percents or
more may be too large to be ignored, however, especially if they
originate from systematic excesses in the equivalent widths of
important radiative lines. Given the link between line equivalent
widths and radiative intensities discussed in \S\ref{sec:EW} in the
context of the narrow-band formalism, we therefore suggest that
equivalent width excesses of the magnitude found in several of our
idealized models could have a significant impact on radiative fluxes
and the overall radiative energy balance of hot Jupiter atmospheres.

However, we must also caution that, beyond these preliminary
arguments, it is not possible to settle this question without a
considerably more detailed calculation than the one presented here.
Indeed, it is the combined effect of bulk Doppler shifts on a vast
number of radiative lines, all with varying degrees of optical
thickness, weighted by the various Planck functions of the atmospheric
layers under consideration (thus depending on the atmospheric
temperature profile itself), along a variety of inhomogeneous optical
paths, which ultimately determines by how much the overall radiative
balance of the dynamic atmosphere is affected. By contrast, our simple
transmissivity and equivalent width models were focused on single
isolated lines with purely linear velocity gradients along arbitrary,
homogeneous optical paths.  Furthermore, contributions from continuum
opacity sources in the atmosphere, for instance in cloudy regions,
could reduce the impact of bulk Doppler shifts on the atmospheric
radiation transport, since small shifts mostly affect narrow radiative
lines. Various simplifying assumptions made in our models would thus
be invalidated in a more realistic radiative transport
calculation. Interestingly, the full radiation transport problem in a
dynamic atmosphere may be amenable to practical numerical solutions in
the future \citep[e.g.,][]{Knop2009}.

It is also worth emphasizing that our discussion has largely focused
on the long-wavelength (thermal) component of the atmospheric
radiation spectrum, for which the diffuse approximation and the use of
an isotropic (Planck) source function are justified. At first, one may
be tempted to neglect the effects of bulk Doppler shifts from
horizontal winds in the treatment of short-wavelength atmospheric
radiation since the corresponding beam-like insolation is vertical in
first approximation. This could be misleading, however. When
scattering is important, as is usually the case, it generates optical
paths slanted enough that they could become much more susceptible to
the bulk Doppler shifts caused by horizontal winds. Furthermore, there
is a significant fraction of the atmosphere that is located far enough
away from the substellar point to be subject to partially slanted
irradiation. In the case of close-in planets like hot Jupiters, the
finite size of the stellar disk could also contribute to slanted
irradiation and thus an increased sensitivity to Doppler shifts from
horizontal winds. Let us emphasize that radiation transport in a
dynamic atmosphere with an anisotropic source function, as would
result from multiple scattering of the beam-like insolation in the
short-wavelength portion of the atmospheric radiation spectrum, would
lead to a considerably more complex formulation of the radiative
transfer problem than discussed here for the case of diffuse thermal
radiation (e.g., Eq.~[\ref{eq:two}]). For example, one wonders whether
the anisotropic transport resulting from atmospheric bulk motions
could result in a stronger polarization signal than has been estimated
on the basis of static radiative transfer calculations
\citep[e.g.,][]{Seager2000}.

Interestingly, the various short-wavelength radiation transport
effects mentioned above should be particularly pronounced in the
context of transmission spectroscopic measurements, which specifically
probe nearly horizontal optical paths at the atmospheric planetary
limb during transits.  Essentially all transit spectroscopic
diagnostics could thus be affected by bulk Doppler shifts from fast
atmospheric winds. \citet{Brown01} has presented a detailed discussion
of this problem, including consequences of the vertical gradients of
horizontal wind velocity. The possibly more important effects of bulk
Doppler shifts along a specific optical path were omitted from these
calculations, however, even though the author did comment on the
expectation of increased line equivalent widths. It may thus prove
important to carefully reassess various interpretations about the
chemical composition of hot Jupiter atmospheres based on transit
spectroscopic measurements with models which properly account for
radiation transport in dynamic, rather than static, atmospheres.

Finally, let us conclude by mentioning a few possible extensions of
this work to other classes of planetary atmospheres. It would seem
that the standard static assumption for the treatment of radiation
transport in Solar System planetary atmospheres is well
justified. Indeed, for these much cooler atmospheres, at a
representative pressure level of 1 bar, radiative lines are
significantly more pressure-broadened in Solar System atmospheres than
in hot Jupiter atmospheres (see scalings in \S\ref{sec:widths}). In
that limit, and for wind speeds well below the sound speed, our models
do indicate very small excesses in the equivalent width of radiative
lines.\footnote{Interestingly, Neptune's and Saturn's upper
  atmospheres and their near-sonic wind speeds
  \citep[e.g.,][]{Suomi1991,Li08} may constitute a partial exception to
  this rule, despite cold atmospheric temperatures.} By contrast, the
typically much hotter planets discovered by astronomers in recent
years, which are often subject to unusually strong radiative forcing
conditions, would seem to be more natural sites for the application of
the radiation-hydrodynamical principles emphasized in the present
work. Besides hot Jupiters, these principles could also find
applications in the class of eccentric giant planets which experience
transient atmospheric flash-heating during periastron passage
\citep[e.g.,][]{Langton2008,Laughlin09} and perhaps the emerging class
of hot super-Earths, if the atmospheres of these planets are able to
sustain wind velocities approaching or exceeding the sound speed in
the presence of significant ground drag.

\acknowledgments

We thank Adam Burrows, Jeremy Goodman, Brad Hansen, Frits Paerels and
Ed Spiegel for useful discussions and an anonymous referee for
comments that helped improve the quality of this manuscript.  This
work was supported by the NASA OSS program, contract \#NNG06GF55G, a
NASA Graduate Student Research Program Fellowship, contract
\#NNX08AT35H, and the Spitzer Space Telescope Program, contract \#
JPLCIT 1366188. We also thank Drake Deming for supporting this work.


\clearpage

\end{document}